\documentclass[doublecol]{epl2} 
\usepackage{amsmath}
\usepackage{amssymb}
\usepackage{graphicx}

\def \be  {\begin{equation}}
\def \ee  {\end{equation}}
\def \ee  {\end{equation}}
\def \bea {\begin{eqnarray}}
\def \eea {\end{eqnarray}}

\title{Phenomenology of strangeness production at high energies}

\author{
Abdel Nasser  Tawfik\inst{1,2} \and Hayam Yassin\inst{3}  \and Eman R. Abo Elyazeed\inst{3} \and Muhammad Maher\inst{2,4}  \and Abdel Magied Diab\inst{1,2} \and Magda Abdel Wahab\inst{3} \and Eiman Abou El Dahab\inst{1,2,5}
} 

\shortauthor{A. Tawfik \etal}

\institute{ 
\inst{1} Egyptian Center for Theoretical Physics (ECTP), MTI University, 11571 Cairo, Egypt. \\
\inst{2} World Laboratory for Cosmology And Particle Physics (WLCAPP), 11571 Cairo, Egypt. \\
\inst{3} Physics Department, Faculty of Women for Arts, Science and Education, Ain Shams University, 11577 Cairo, Egypt. \\
\inst{4} Physics Department, Faculty of Science, Helwan University, 11795 Cairo, Egypt. \\
\inst{5} Faculty of Computer Sciences, MTI University, 11671 Cairo, Egypt
}

 {Relativistic heavy-ion nuclear reactions}
 {Particle production (relativistic collisions)}
 {Quantum chromodynamics in nuclei}

\abstract{
The strange-quark occupation factor ($\gamma_s$) is determined from the statistical fit of the multiplicity ratio $\mathrm{K}^+/\pi^+$ in a wide range of nucleon-nucleon center-of-mass energies ($\sqrt{s_{NN}}$). From this single-strange-quark-subsystem, $\gamma_s(\sqrt{s_{NN}})$ was parametrized as a damped trigonometric functionality and successfully implemented to the hadron resonance gas model, at chemical semi-equilibrium. Various particle ratios including $\mathrm{K}^-/\pi^-$, $\mathrm{\Lambda}/\pi^-$, and $\mathrm{\bar{\Lambda}}/\pi^-$ are well reproduced. The phenomenology of $\gamma_s(\sqrt{s_{NN}})$ suggests that,  the hadrons ($\gamma_s$ raises) at $\sqrt{s_{NN}} \simeq 7~$GeV  seems to undergo a phase transition to a mixed phase ($\gamma_s$ declines), which is then derived into partons ($\gamma_s$ remains unchanged with increasing $\sqrt{s_{NN}}$), at  $\sqrt{s_{NN}} \simeq 20~$GeV.
}

\begin{document}

\maketitle

%--------------------------------------------------------------------------------
%                                        Introductions
%--------------------------------------------------------------------------------
\section{Introduction  \label{sec:intro} }

The energy scan of the Super Protonsynchrotron (SPS) confirming continuation and completion of the set of excitation functions observed at the Alternating Gradiant Synchrotron (AGS)  - among others - the ratios of various particle multiplicities. Widely-known examples include the so-called Marek's {\it  horn-like} structure of  $\mathrm{K}^+/\pi^+$ \cite{Marek1999}, which was confirmed in various heavy-ion collisions experiments \cite{Expp1,Expp2,Expp3}, the {\it kick}, which is to be drawn through a rapid change in the slope of thr produced pions per participating nucleon and the {\it plateau} of the averaged transverse momentum $\langle p_T\rangle$ of kaons, known as {\it step}. The recent energy scan of the Relativistic Heavy Ion Collider (RHIC) confirms the horn-like structure and both kick and step observations \cite{Expp2}, as well. These phenomena apparently indicate remarkable changes taking place in the strongly interacting system at relativistic energies. Various explanations were proposed, so far \cite{Marek1999,Reff14,Reff16,Reff13,Expln1,Expln1b,Expln2,Expln3}, including the
onset of deconfinement, the manifestation of critical endpoint, etc. That the thermal models seem not being able to reproduce such a remarkable structure represents one of the chronic puzzles facing  the particle scientists, to which a great number of experiments and theoretical and numerical studies were devoted.

When focusing the discussion of the horn-like structure of $\mathrm{K}^+/\pi^+$, we first observe an increase in $\mathrm{K}^+/\pi^+$ with increasing the
nucleon-nucleon center-of-mass energy ($\sqrt{s_{NN}}$) up to a certain
value (at projectile energy of $\sim 30~$AGeV \cite{Expp4}), which is then followed by a rapid decrease.  At higher $\sqrt{s_{NN}}$, $\mathrm{K}^+/\pi^+$ remains almost
unchanged shaping a plateau at top RHIC energies. This continues up the Large hadron Collider (LHC) energies, where another puzzle was reported, recently, namely the proton anomaly or the discrepancy between calculated and measured $\mathrm{p}/\mathrm{\pi}$ ratio \cite{ppi}. It was conjectured that, this energy-dependence can be generated by the hadronic kinetic model \cite{Reff12} and also when assuming a transition from a baryon-dominated system (at low) to a meson-dominated system (at high collision energy) \cite{Reff14}. Introducing heavy masses of {\it unknown} hadronic resonances (Hagedorn mass spectrum) \cite{Reff13} is believed to assure a release of additional color degrees-of-freedom (dof) \cite{Marek1999} so that large numbers of pions even greater than that of the kaons can be guaranteed. The latter - in turn - can be achieved when including heavy hadron resonances from the recent compilation of the Particle Data Group (PDG) \cite{Reff16} in the HRG model. All these proposals were implemented but unfortunately not able to give an unambiguous clarification, so far. 

For the production and evolution of strange quarks (and strange hadrons) with varying collision energies, a microscopic model was alternatively utilized \cite{Reff}, where the momentum integrated Boltzmann equation should be first evaluated. In the Bag model, the ratios of strangeness to entropy was studied in dependence on the collision energy $\sqrt{s_{NN}}$ \cite{Rref5}.  Interestingly, a scaling behavior similar to the one from hadronic nonequilibrium kinetic model, which considers the energy-depending lifetime of the fireball \cite{{Expln1b}} was obtained. Both are found similar to the observed $\mathrm{K}^+/\pi^+$. The present letter introduces an energy-dependent strange-quark occupation factor ($\gamma_s$) deduced, phenomenologically, and assures a best reproduction of various particle ratios.

The  occupation factors of light- and strange-quarks, $\gamma_l$ and $\gamma_s$, respectively, were first introduced by Letessier and Rafelski as a plausible explanation for the strangeness enhancement \cite{Rgmm1}, which was proposed as a sensitive signature for the formation of quark-gluon plasma (QGP). It was argued that the energy threshold for $\bar{s}s$ pair-production considerably differs due to the underlying QCD symmetries. In QGP phase, $\epsilon_{\mathrm{thr}}\simeq 300~$MeV, while in the hadron phase, $\epsilon_{\mathrm{thr}}\simeq 700~$MeV, because the energy threshold corresponds to two times the rest mass of partons and hadrons of interest. Thus, it was conjectured that the creation of QGP should be accompanied by an increase in the strangeness production. This was first confirmed, experimentally \cite{Rref3} and also in the first-principle lattice QCD simulations \cite{Rref4}. Also, it was assumed that the nonequilibrium values, i.e. the ones differing from unity, assigned to $\gamma_s$ would be able to characterize the {\it legend} energy-dependence of $\mathrm{K}^+/\pi^+$ \cite{Expln2}. The present work is an extension and updating of Ref. \cite{Expln2} with most recent experimental results and more plausible phenomenological explanations. 

From the experimental point-of-view, the relative production of strange and nonstrange quarks was analysed  in elementary and in nucleus collisions \cite{Rref1}. Relative the the light-quark pair production, it was concluded that $\bar{s}s$ pairs would be considerably suppressed. Through quark counting, one assumes that $\bar{u}u:\bar{d}d:\bar{s}s=1:1:\lambda_s$, where $\lambda_s\equiv \gamma_s\ll 1$ apart from some factors characterizing the results from heavy-ion collisions. This observation was first confirmed in heavy-ion collisions at $\sqrt{s}\simeq 60~$GeV \cite{Rref2}. It was reported that the experimental estimation for average multiplicities of $\langle N_{\bar{s}s}\rangle$ allowed the conclusion that $\lambda_s$ is not sensitive to the interacting system, e.g. quantum numbers of colliding nuclei, but apparently to the collision energies \cite{Rref2}. 
\bea
\lambda_s &=& \frac{\langle N_{\bar{s}s}\rangle}{\langle N_{\bar{q}q}\rangle-\langle N_{\bar{s}s}\rangle},
\eea
where $\langle N_{\bar{q}q}\rangle=\langle N_{\bar{u}u}\rangle+\langle N_{\bar{d}d}\rangle+\langle N_{\bar{s}s}\rangle$, i. e. $\lambda_s$ can be determined from the detection of strange mesons and the total meson multiplicity.

\section{Approach}

In the partition function of a grand-canonical ensemble describing a strongly
interacting system, statistically, $\gamma_l$ and $\gamma_s$ can be integrated
in as pre-factors to the Boltzmann exponential,
\bea
{\ln Z}\left(T,\mu\right) &=&\pm \sum_i \frac{Vg_i}{2{\pi }^2}  \int^{\infty }_0
k^2\, dk \nonumber \\
&& \ln  \left[1 \pm \left(\gamma_l^{n_l}\right)_i\, \left(\gamma_s^{n_s}\right)_i \, 
  e^{\left(\frac{\mu_i-\varepsilon_{i}(k)}{T}\right)}\right], \label{eq:lnZ} 
\eea
where $\pm$ stands for fermions and bosons, respectively, and $\varepsilon
_{i}(k) =(k^{2} +m_{i}^{2})^{1/2}$ is the energy-momentum relation of the $i$-th
hadron. Implementing Hagerdon picture that heavy resonances are composed of
lighter ones which in tern are consisting of ligher ones and so on, the
subscript $i$ refers to a summation over fermions and bosons from recent compilation of PDG. The chemical potential $\mu_i$ is composed of various contributions, for instance, 
\begin{equation} 
\mu_i=B_{i} \mu _{B} + S_{i} \mu _{S} + \cdots,
\end{equation}
where $B_i$ and $S_i$  are baryon and strangeness quantum number of $i$-th
hadron resonance and  $\mu_{B}$ and $\mu_{S}$ are the baryon and strangeness
chemical potential, respectively. $V$ and $T$ are the fireball volume and
temperature, respectively. 

In Eq. (\ref{eq:lnZ}), $n_l$ and $n_s$ are the number of light- and strange-quarks, respectively, of which each fermion or boson is composed. When unity is assigned to $\gamma_l$ and $\gamma_s$, full chemical equilibrium, as the case in the {\it equilibrium} hadron resonance gas (HRG) model, the calculated $\mathrm{K}^+/\pi^+$ poorly generates the horn-like structure measured at  SPS energies, but greatly overestimates the results at high $\sqrt{s_{NN}}$. It is noteworthy highlighting that at AGS energies, the {\it equilibrium} HRG calculations seem to describe well the production rates of strange hadrons \cite{Expln2,Expln3,Satz2016}. In this energy-limit, the HRG calculations seem not sensitive to $\gamma_s$ as to the resonance masses and the excluded volume corrections, etc., Fig. \ref{fig:2}.  The statistical nature of strangeness production, in our case kaon, is conjectured to be maintained, when {\it ad hoc} nonequilibrium values are assigned to the strange-quark occupation factor ($\gamma_s\neq1$) \cite{Rgmm1}. The present letter concludes a rapid strangeness enhancement at AGS energies, at equilibrium light-quark occupation factor ($\gamma_l=1$), i.e. chemical semi-equilibrium.

A strangeness suppression factor, i.e. $\gamma_s<1$, was explicitly assumed in Ref. \cite{Rgmm2}, 
\bea
\gamma_s(\sqrt{s_{NN}}) &=& 1 - \alpha \, \exp\left(-\beta\, \sqrt{A\sqrt{s_{NN}}}\right), \label{eq:gammas}
\eea
where $\alpha=0.606$, $\beta=0.0209$, and $A$ is the atomic numbers of the colliding nuclei. This has the advantage to reduce the production rate of the strange hadrons \cite{Rgmm2}.

In the present work, we also assume $\gamma_l=1$. But for $\gamma_s$, we fit our HRG calculations, where the number density can be derived from Eq. (\ref{eq:lnZ}), on $\mathrm{K}^+/\pi^+$ at varying $\sqrt{s_{NN}}$ with the experimental results, top panel of Fig. \ref{fig:1}. From the statistical fit to this single-strange-quark-subsystem, we aim to parametrize $\gamma_s$ as functionality of $\sqrt{s_{NN}}$. Details about HRG can be taken from Ref.  \cite{Tawfik:2014eba}, where $\sqrt{s_{NN}}$ is related to baryon chemical potential ($\mu_B$) and thus directly enters the partition function, Eq. (\ref{eq:lnZ}). For specific particle species, their number densities are composed of the contributions coming from the corresponding hadrons and that stemming from heavy resonances decaying into that particles of interest. The latter should be weighted by the corresponding branching ratios.

Within the extensive statistics, it is conjectured that, the colliding system reaches the stage of chemical freezeout, which is characterized by two thermodynamic quantities; $T$ and $\mu_B$, which can be deduced from statistical fits of the HRG calculations of various particle ratios and/or yields to the experimental multiplicities. Resulting $T$ and $\mu_B$ are well described by various freezeout  conditions \cite{Tawfik:2014eba,Tawfik:2015kwa,Tawfik:2016jzk,Tawfik:2014dha}. Concretely, we implement $s/T^3=7$ \cite{Tawfik:2005qn,Tawfik:2004ss}, with $s$ being the entropy density. In fitting the experimental $\mathrm{K}^+/\pi^+$ to the HRG calculations, we assume $\gamma_s$ as a free parameter, while the values of the parameters $T$ and $\mu_B$ are determined at $s/T^3=7$ and $\gamma_q$ is assigned to a constant value. As mentioned, we concentrate the discussion on fitting with the single-strange-quark-subsystem $\mathrm{K}^+/\pi^+$ and assume that the resulting $\gamma_s(\mathrm{K}^+/\pi^+)$ remains valid to the entire HRG thermal model.

\section{Results}

\begin{figure}[htb]
\centering
\includegraphics[width=5.cm,angle=-90]{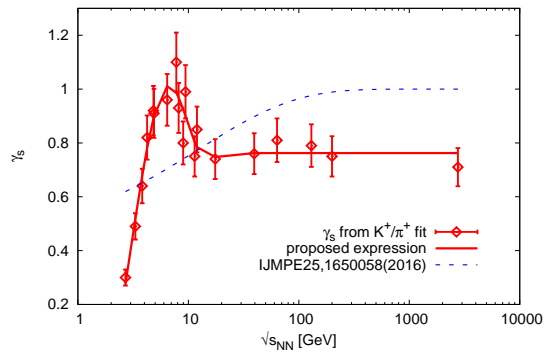} 
\caption{Strange-quark occupation factor, $\gamma_s$, is given as a function of $\sqrt{s_{NN}}$. Symbols with errors depict our results from the statistical fits of measured $\mathrm{K}^+/\pi^+$ to the HRG calculations with $\gamma_s$ taken as a free parameter. The dashed curve represents Eq. (\ref{eq:gammas}). 
\label{fig:1}}
\end{figure}

\begin{figure}[htb]
\centering
\includegraphics[width=4.5cm,angle=-90]{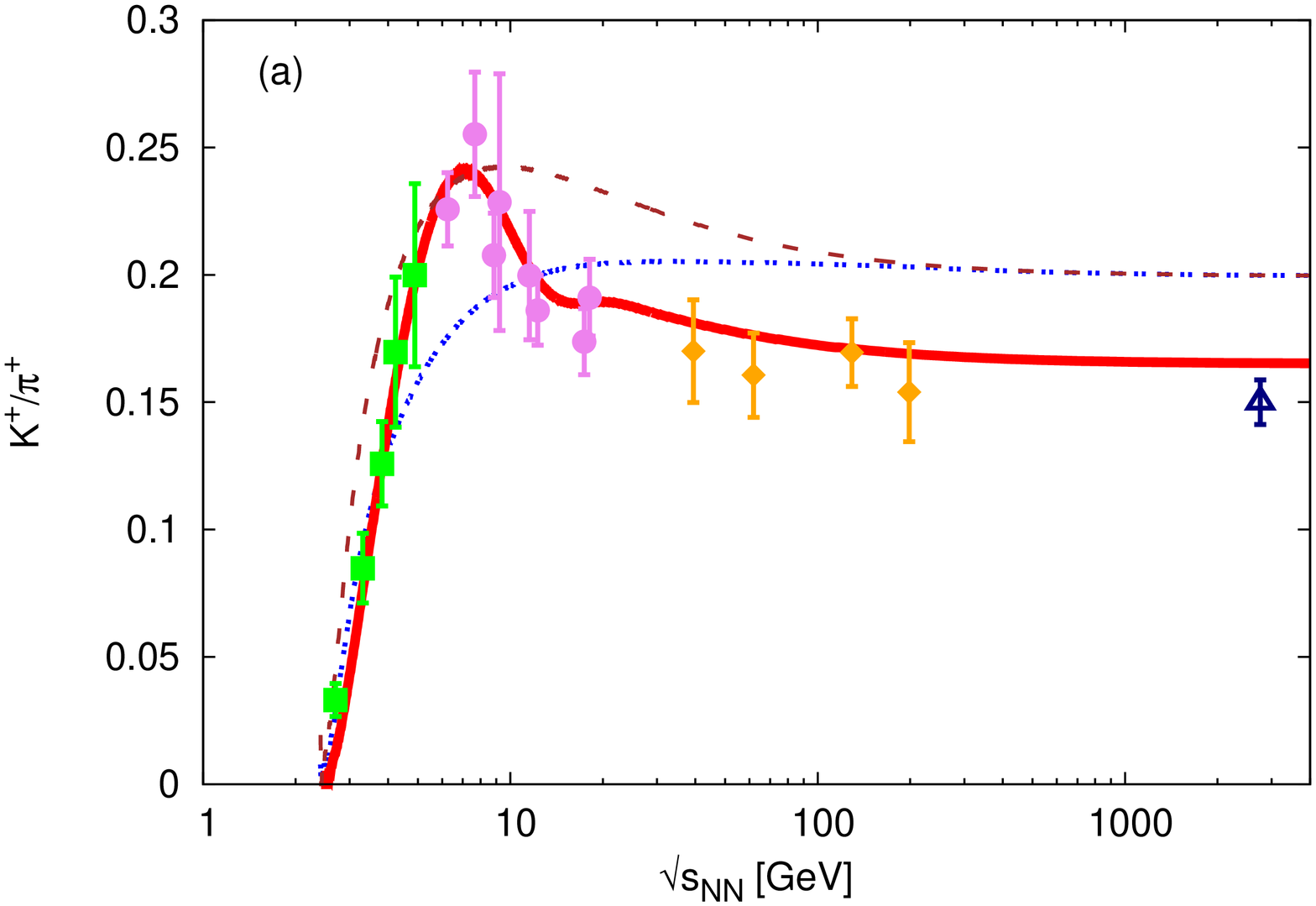} \\
\includegraphics[width=4.5cm,angle=-90]{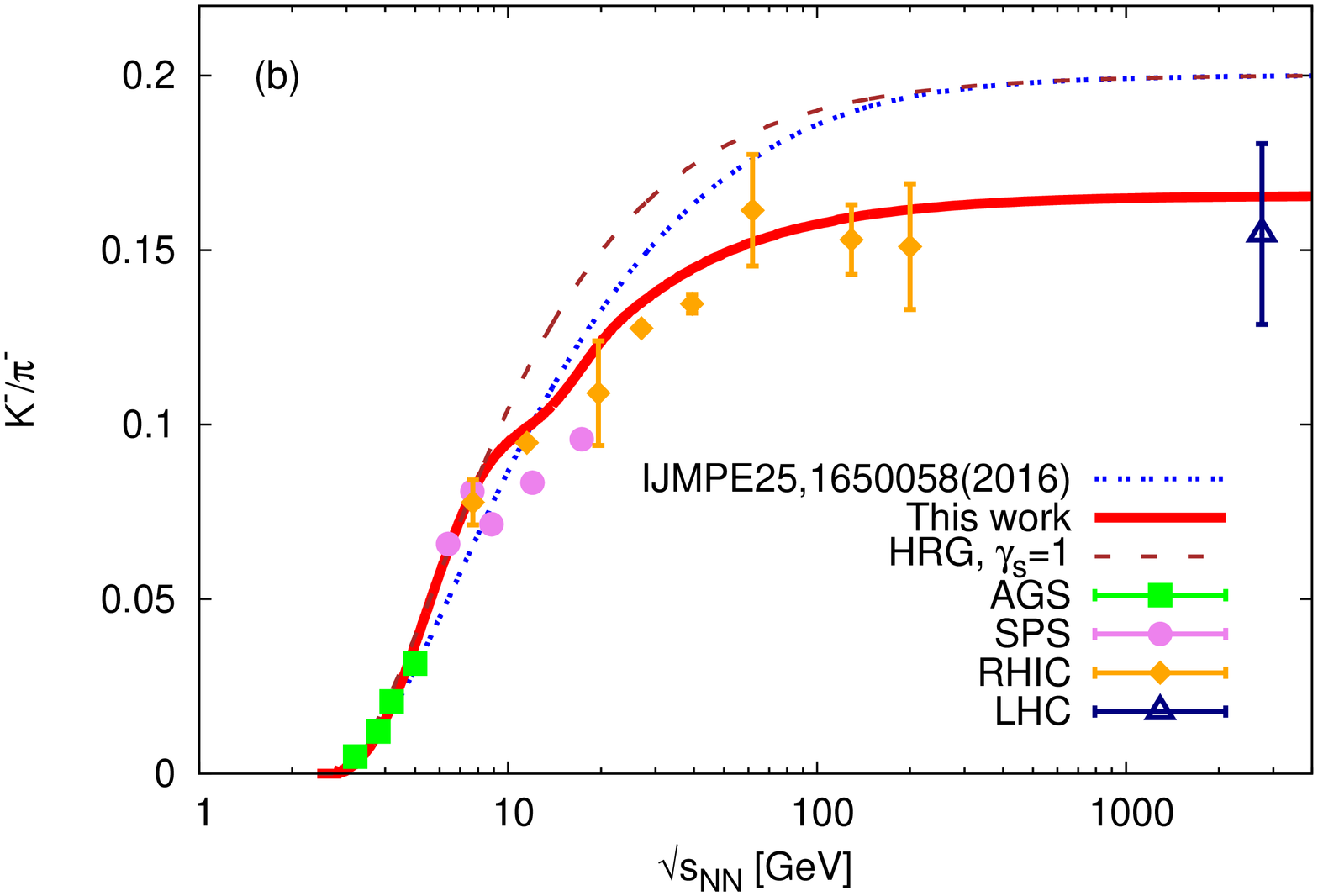} \\
\includegraphics[width=4.5cm,angle=-90]{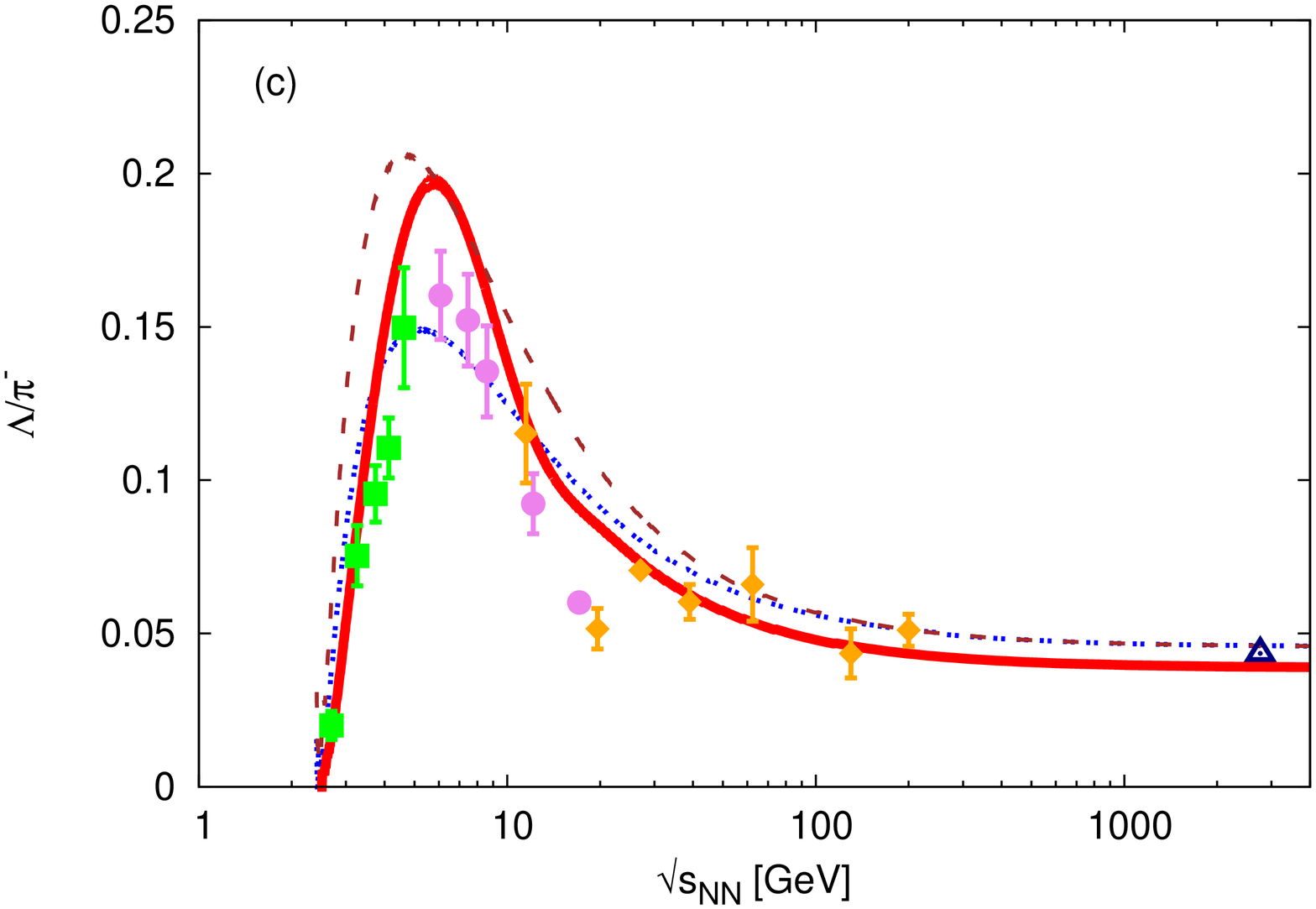}\\
\includegraphics[width=4.5cm,angle=-90]{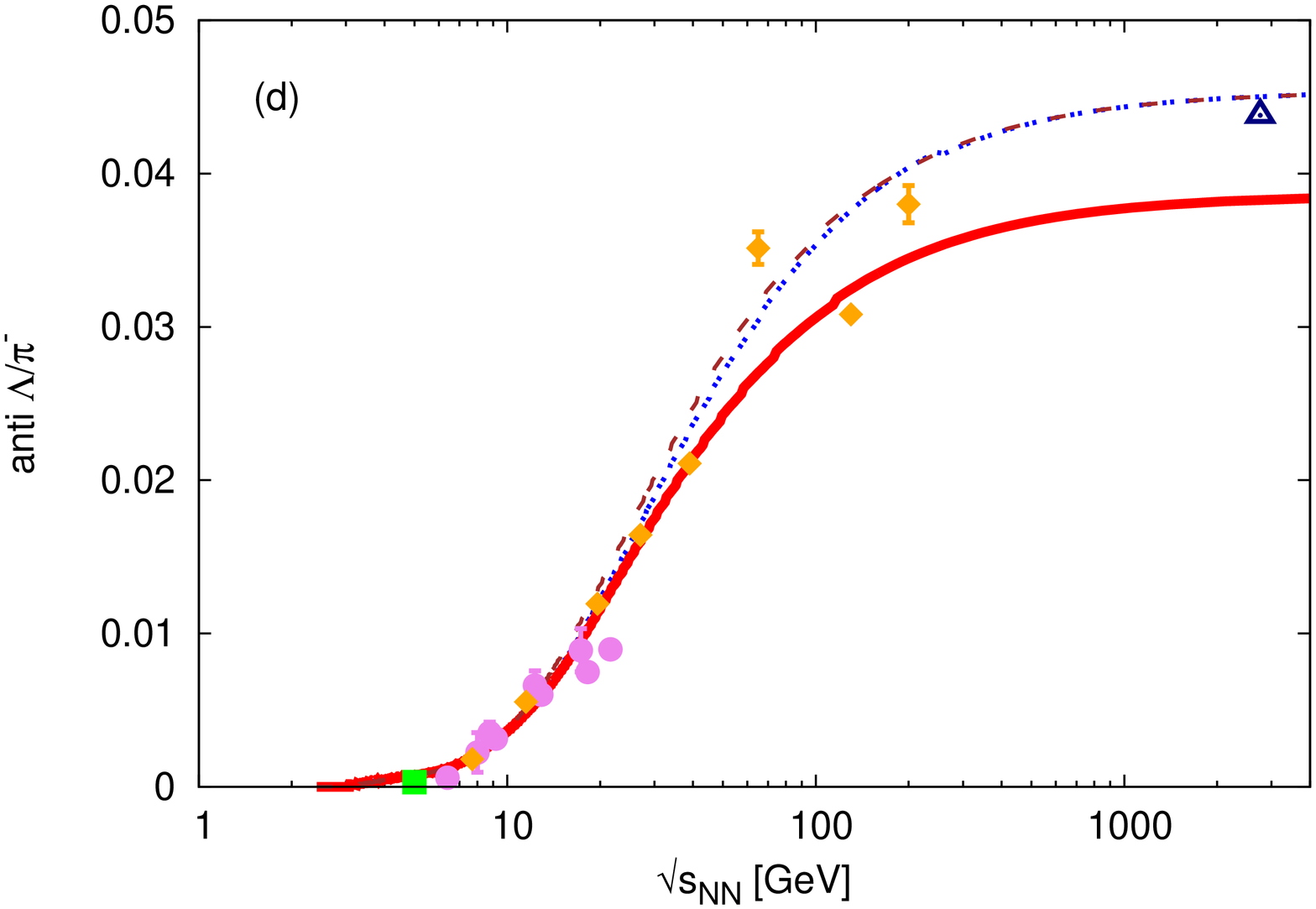}
\caption{The experimental results of different particle-ratios are given in dependence on  $\sqrt{s_{NN}}$. The solid curves depict the present calculations at $\gamma_l=1$ and $\gamma_s$ is determined from  Eq. (\ref{eq:ourFit}). Double-dashed and dashed curves represent {\it ideal} HRG at full chemical equilibrium ($\gamma_l=\gamma_s=1$) and  semi-equilibrium [$\gamma_l=1$ and $\gamma_s$ is given by  Eq. (\ref{eq:gammas})], respectively. }
\label{fig:2}
\end{figure}

In Fig. \ref{fig:1}, the resulting $\gamma_s$ are depicted as a function of $\sqrt{s_{NN}}$ (symbols with errors). At the chemical freezeout, these are well described by a damped trigonometric functionality, 
\bea
\gamma_s(\sqrt{s_{NN}}) &=& a\, \exp\left(-b\, \sqrt{s_{NN}}\right)\, \sin\left(c\, \sqrt{s_{NN}} + d\right) \nonumber \\
&+& f, \label{eq:ourFit} 
\eea
with $a=2.071\pm0.259$, $b=0.282\pm 0.062$, $c=0.362\pm0.051$, $d=4.78\pm0.21$, and $f=0.764\pm0.249$. Expression (\ref{eq:ourFit}) is illustrated as the solid curve. For the sake of completeness, we compare our results with predictions deduced from Eq. (\ref{eq:gammas}), which is given as the dashed curve. So far, both expressions do not allow for any concrete conclusion. This might be only possible, when they are (or not) succeeded in reproducing other particle ratios as illustrated in Fig. \ref{fig:2}.  

When the energy-dependence of $\gamma_s$, Eq. (\ref{eq:gammas}), \cite{Rgmm2} is implemented, we first find that the resulting $\gamma_s$ is monotonically depending on $\sqrt{s_{NN}}$.  At $\sqrt{s_{NN}}\lesssim 200~$GeV, we find that, $\gamma_s$ exponentially increases, while at higher energies, $\sqrt{s_{NN}}$ likely approaches an asymptotic value; the chemical equilibrium. At $\gamma_l=1$ and Eq. (\ref{eq:gammas}), four particle ratios are calculated and shown on Fig. \ref{eq:gammas}. This shall be elaborated, shortly. For now, we merely highlight that, the results are almost identical to the ones calculated at full chemical equilibrium. Accordingly, one can easily judge about the improvement made through Eq. (\ref{eq:gammas}).

Our parametrization for $\gamma_s$, Eq. (\ref{eq:ourFit}), contrarily shows a nonmonotonic energy-dependence. At $\sqrt{s_{NN}}\lesssim 7~$GeV, $\gamma_s$ exponentially increases with increasing $\sqrt{s_{NN}}$. Then, within the energy range $7 \lesssim \sqrt{s_{NN}}\lesssim 20~$GeV, $\gamma_s$ rapidly decreases. At higher energies, $\gamma_s$ becomes energy independent, especially at top RHIC and LHC energies. Its asymptotic value is given by the parameter $f$ in Eq. (\ref{eq:ourFit}).

The top panel of Fig. \ref{fig:2} shows $\mathrm{K}^+/\pi^+$ (symbols) as a function of $\sqrt{s_{NN}}$, which in turn is related to $\mu_B$. Dashed curve gives HRG calculations, at equilibrium occupation parameters, namely full chemical equilibrium $\gamma_l=\gamma_s=1$. The double-dotted curve illustrates results at $\gamma_l=1$, while $\gamma_s$ is determined from Eq. (\ref{eq:gammas}) \cite{Rgmm1}. The earlier estimations excellently reproduce the $\mathrm{K}^+/\pi^+$ results at $\sqrt{s_{NN}}\lesssim 10~$GeV. At higher energies, although they overestimate the experimental results, a horn-like structure remains guessable. The latter calculations don't posses any horn-like structure. At high energies, they fit well with the results from equilibrium HRG, i.e. overestimate the measurements, as well. 

Our results are presented as the solid curve. They are deduced from {\it ideal} HRG calculations, i.e. point-like constituents, PDG compilation, etc. In these calculations, chemical semi-equilibrium, namely $\gamma_l(\sqrt{s_{NN}})=1$ but $\gamma_s(\sqrt{s_{NN}})$ is to be determined from Eq. (\ref{eq:ourFit}) are assumed. From the fact that Eq. (\ref{eq:ourFit}) is stemming from the statistical fit of the HRG calculations to the measured $\mathrm{K}^+/\pi^+$, the excellent agreement between the solid curve and $\mathrm{K}^+/\pi^+$ results is very obvious. To judge about the validity of our parametrization,  Eq. (\ref{eq:ourFit}), we still need to utilize it in calculating other quantities and we might also need to run further checks, such as, lattice QCD thermodynamics and examine the thermodynamic consistency, etc. The latter shall be subjects of future studies. For the phenomenological focus of the present letter, we concentrate on calculating various particle ratios. 

Panels b), c), and d) of Fig. \ref{fig:2} depict $\mathrm{K}^-/\pi^-$, $\mathrm{\Lambda}/\pi^-$, and $\mathrm{\bar{\Lambda}}/\pi^-$, respectively, as functions of $\sqrt{s_{NN}}$. These experimental results are confronted to the HRG calculations at full chemical equilibrium, i.e. $\gamma_l(\sqrt{s_{NN}})=\gamma_s(\sqrt{s_{NN}})=1$ (dashed curves) and quasi-equilibrium, i.e. $\gamma_l(\sqrt{s_{NN}})=1$ and $\gamma_s(\sqrt{s_{NN}})=1$. The latter is given by  Eq. (\ref{eq:gammas}) \cite{Rgmm1} (double-dotted curves). We find that they are almost compatible with the HRG calculations at full chemical equilibrium. 

Our HRG calculations with Eq. (\ref{eq:ourFit}) are presented as the solid curves. Here, we find that, the agreement with the experimental results is remarkably improved. Thus, we conclude that our parametrization for $\gamma_s(\sqrt{s_{NN}})$ enables the thermal models, the HRG for instance, to reproduce various particle ratios. Relative to Eq. (\ref{eq:gammas}), Eq. (\ref{eq:ourFit}) agrees very well with the experimental results.

\section{Discussion and Conclusions}
\label{sec:disc}

As discussed in Introduction, various interpretations for the {\it horn-like} structure measured in $\mathrm{K}^+/\pi^+$ have been proposed, for example, parametrization for the evolution of the fireball and the produced-particle densities \cite{Expln1b}. It was noticed that, for beam energies $>30~$AGeV, the fireball lifetime decreases, whose interplay with the initial energy density determines the final-state density. The hadronic kinetic model based on these assumptions besides full chemical equilibrium are conjectured to describe well different particle ratios including $\mathrm{K}^+/\pi^+$.

Also, to interpret the energy-dependence of the kaon production, a microscopic approach and full chemical equilibrium have been proposed \cite{Reff}. Here, the evolution of strangeness production was modelled by momentum integrated Boltzmann equation. This approach was borrowed from the freezeout processes in the thermal expansion of the early universe \cite{Rref10}. When an initial partonic phase is assumed at collision energies greater than a certain threshold, a non-monotonic energy-dependence of $\mathrm{K}^+/\pi^+$ was obtained. 

The excitation functions in the particle multiplicities $\mathrm{K}^+/\pi^+$ observed at SPS energies and confirmed in the RHIC energy-scan, especially the {\it horn-like} structure, are analysed in the HRG statistical-thermal model, in which point-like constituents and chemical semi-equilibrium are considered. The strange-quark occupation factor is assumed to vary with the collision energy. This idea was introduced and worked out by many authors, including AT. Additional to a best reproduction of various particle ratios, the present work introduces an updating with most recent experimental results and presents a novel parametrization of the dependence of $\gamma_s$ on the collision energies. The resulting functionality describing $\gamma_s$ with varying $\sqrt{s_{NN}}$ is proposed as damped trigonometric functionality. Accordingly, such a monotonic energy-dependence of $\gamma_s$ was successfully implemented to the HRG model at chemical semi-equilibrium. As discussed, the particle ratios $\mathrm{K}^-/\pi^-$, $\mathrm{\Lambda}/\pi^-$, and $\mathrm{\bar{\Lambda}}/\pi^-$ are well reproduced. 

In light of this phenomenologic well description and the assumption that the light-quark occupation factor is assigned to the equilibrium value, we can conclude that $\gamma_s\lesssim 1$, i.e. chemical semi-equilibrium, at all collision energies. Only at low SPS energy, one obtains that $\gamma_s\rightarrow 1$ and even slightly exceeds this characteristic value deriving the colliding system to full chemical equilibrium. Accordingly, we expect that such a state should possess maximum entropy. Although, other aspects on thermodynamic consistency shall be analysed in future works, we merely highlight that an old paper of AT \cite{Expln2} reported on the energy-dependence of $s/n$, where $n$ and $s$ being number and entropy densities, respectively. It was concluded that the {\it horn-like} structure of $\mathrm{K}^-/\pi^-$ is positioned where maximum $s/n$ takes place.

So far, we conclude that, at $\sqrt{s_{NN}}\lesssim 7~$GeV, the strongly interacting system is characterized by strangeness enhancement, i.e. $\gamma_s$ exponentially increases. Within this energy region, symmetry and dof seem to characterize hadron matter. After reaching a maximum value, $\gamma_s\rightarrow 1$, another symmetry and other effective dof come to play an increasing role. This leads to a rapid decrease in $\gamma_s$.  Within $7 \lesssim \sqrt{s_{NN}}\lesssim 20~$GeV, the value of $\gamma_s$ decreases from $\sim 1$ to $\sim 0.76$. It is believed that the hadron system is derived into a mixed phase, where both hadronic and partonic dof and symmetries exist, simultaneously.  At higher energies, $\gamma_s$ reaches its asymptotic value, $f\simeq 0.76$. The interacting system is conjectured to undergo another phase transition into a partonic phase. 

With this phenomenological description, we emphasize how $\gamma_s$ varies with the collision energy and implement this in calculating various particle ratios. We don't argue that $\gamma_s$ plays the role of an order parameter. Much more, it is a quantity manifesting imprints of the QCD phase transition. Apparently, its value considerably changes with changing the underlying dof and symmetries. The latter could be utilized as order parameters. Although, its energy-dependence points to first-order one going through an inhomogeneous phase composed of two states in a chemical quasi-equilibrium, the order of the phase transition itself wouldn't be conveyed though $\gamma_s$, merely.

As introduced, we propose the functionality $\gamma_s(\sqrt{s_{NN}})$ deduced from the statistical fit to the single-strange-quark-subsystem $\mathrm{K}^+/\pi^+$ and assume to be applicable to the entire HRG model. This raises the question whether multi-strange-quark-subsystems such as $\phi/\pi$ or $\Xi/\pi$ and $\Omega/\pi$, would revise this phenomenological picture? Another future work shall be devoted to formulate answers to this interesting question.

%%%%%%%%%%%%%%%%%%%%%%%%%%%%%%%%%%%%%%%%
%%%             references
%%%%%%%%%%%%%%%%%%%%%%%%%%%%%%%%%%%%%%%%

\end{document}